\documentclass{article}
\usepackage{spconf,amsmath,graphicx}
\usepackage{bbold}
\usepackage{xcolor}

\usepackage{booktabs,multirow}
\usepackage{bigdelim}
\usepackage{float}

\usepackage{algorithm}
\usepackage{algorithmic}

\usepackage{hyperref}
\usepackage{cleveref}
\usepackage{graphicx}
\usepackage{multirow}

\title{2-Dimensional spectral gating for denoising bioacoustics recordings}
\name{Julien Boussard$^{1,3\hspace{0.15em}}$ \qquad Mélisande Teng$^{1,2\hspace{0.15em}}$ \qquad Sulagna Saha$^{1,3}$ \qquad Mario Gallego-Abenza$^{4}$}
  
  \address{$^{1}$ Mila - Quebec AI Institute,
  $^{2}$ Pioneer Centre for AI, University of Copenhagen,
      $^{3}$ McGill University, \\ $^{4}$ Stockholm University
}
\begin{document}
%
\maketitle
\begin{abstract}
Isolating vocalizations from noise in bioacoustics recordings is a prerequisite to many ecological analyses, including species identification, animal communication understanding, and individual or population-level variability studies. However, when recordings are acquired in open environments, vocalizations, noise, or signal-to-noise ratio can vary widely across individuals, species, environment, and recording conditions, making it hard to develop robust and generalizable methods for ecological analyses. To account for these challenges, noise reduction techniques are used to remove noise before downstream analyses. Popular methods such as Noisereduce rely on spectral gating, which estimates a noise threshold for each frequency channel. We propose a further improvement to Noisereduce, leveraging the fact that most animal sounds have structure across multiple frequencies. We apply our method on bird and marine mammals recordings and show that our extension of Noisereduce leads to improved denoising in both above and under water acoustic recordings, without impacting speed of preprocessing.

\end{abstract}
\section{Introduction}

Bioacoustics, the study of animal sounds, has emerged as an important tool for conservation \cite{PENAR2020100847}. Recently, vast amounts of bioacoustics recordings have become increasingly available thanks to advances in recording technologies and citizen science initiatives \cite{xeno_canto}. These recordings are acquired in heterogeneous conditions, far from controlled environments such as soundproof chambers, and present a wide variation in types of noise, signal-to-noise ratio, and vocalizations. This poses challenges for downstream ecological analyses of these recordings, which often rely on accurate signal extraction or detection \cite{noise_challenges}. 
Reliable denoising is also important in a context where machine learning (ML) approaches leveraging large amounts of data open promising avenues for bioacoustics but can fail at generalizing across recording conditions and vocalization types or overfit to background noise instead of extracting useful information \cite{background_noise_ml}.  
This highlights the need for robust denoising methods that effectively remove the noise, regardless of the recording environment. 


Traditional denoising approaches are mostly filters. Bandpass filters \cite{bandpass_filter} mask out low or high frequencies globally to isolate frequencies where the signal is present. Other filters such as the mean, median or Gaussian filters smooth the signal by pulling the mean, median, or weighted mean, over a sliding window. The Wiener filter \cite{wiener1949}, widely used in acoustics, assumes stationary signal and additive noise to estimate the signal by linear time-invariant filtering, and the Savitzky-Golay filter \cite{savitzky_golay} fits a low-degree polynomial to a local sliding window using linear least squares to estimate the signal. While these methods are fast and efficient, they generally assume additive, Gaussian noise and generalize poorly to different types of natural noise. 

To remedy these issues, spectral subtraction or gating approaches estimate the noise per frequency band. They first transform the signal into the time-frequency domain, estimate the noise spectrum, and either subtract it from the raw spectrum or mask out noise frequencies. In particular, Noisereduce \cite{noise_reduce} works as follows. It first estimates the mean and standard deviation ($\mu_f, \sigma_f$) for each frequency band $f$, and then masks every value that is below $\mu_f + z_{\text{th}} * \sigma_f$, to keep all the signal that is considered an outlier under the noise distribution. $z_\text{th}$ represents a z-score. 
By treating each frequency band independently, this method does not leverage signal structure.  

More recently, several works have proposed deep-learning based denoising approaches that train an autoencoder to denoise the signal by mapping it to a low dimensional space \cite{AutoencoderDenoiser1},
train a supervised neural network using manual annotations \cite{amine_denoiser}, or use a noise model to generate pairs of noisy and clean samples and train a supervised denoising model \cite{noise2noise}. These approaches require large training datasets and machine learning expertise, labelled data, or a good noise model, hindering their generalization capabilities and their adoption by the bioacoustics community. 

In this paper, we propose an extension to Noisereduce. Our 2D spectral gating method (2D-SG) leverages the structure of animal vocalizations, which usually exhibit signals that span across multiple frequencies. We do not compare our approach to deep learning methods in this paper, as our goal is to propose a simple, efficient method that is readily applicable to any recording, and requires neither machine learning expertise nor additional data labeling or collection.


\section{Method}\label{sec:method}

\begin{figure}[htbp]
    \centering
    \includegraphics[width=\columnwidth]{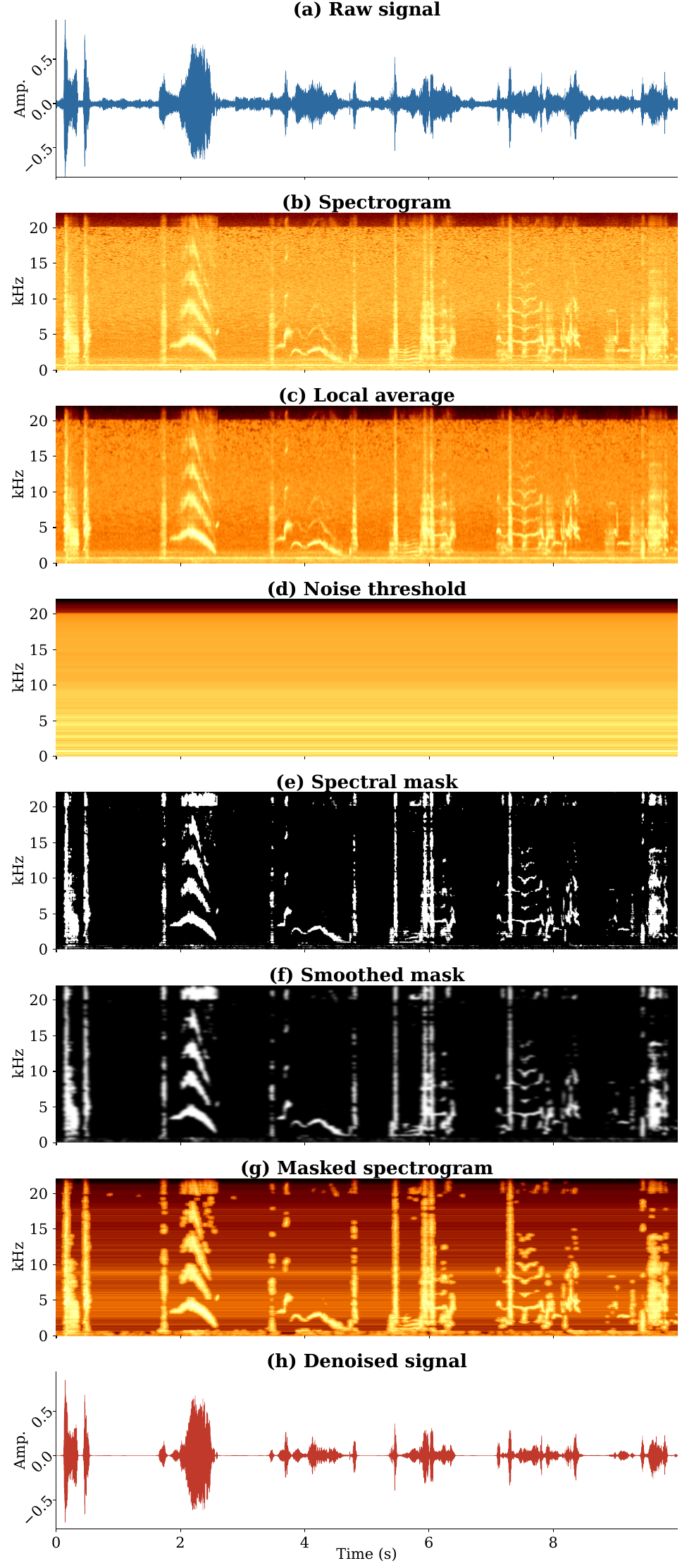} 
    \caption{\textbf{Illustration of the different steps of our 2-dimensional spectral gating method (2D-SG).} The raw audio signal (a) is converted to a spectrogram (b), which is then smoothed using a Gaussian filter to compute a local average (c). The noise threshold (d) is estimated from the raw spectrogram and the mask (e, f) is applied to the spectrogram (g) before inverting it to reconstruct the denoised signal (h). The figure uses the ``B335'' recording from the NOIZEUS Birdsong dataset, with SNR 10 dB and ``waterfall'' noise.}
    \label{fig:method}
\end{figure}

2D-SG is described in Algorithm \ref{alg:augmented_nr} and illustrated in Fig. \ref{fig:method}. Similar to Noisereduce, it takes as input an audio recording (a) and computes a spectrogram (b). Then, it runs a Gaussian filter over the spectrogram to estimate a 2D local average (c). This filter smooths out noise outliers and increases the gain where signal is shared across frequencies. This is suited for animal vocalizations which generally exhibit structure over multiple frequencies. The size of the Gaussian filter window can be chosen wider on the frequency axis than on the time axis for animals that have short vocalizations (e.g. clicks). Then, it estimates the noise distribution and threshold per channel (d) and computes a mask where the local average is lower than the threshold (e). It then optionally smooths the mask (f) to avoid spectral leakage, discontinuity, and sampling artefacts such as the Gibbs phenomenon \cite{jerri1998gibbs}, and applies it before inverting the Fourier transform to reconstruct the denoised signal. 

2D-SG differs from Noisereduce as it computes the mask from the 2D local average of the spectrogram (c) rather than the spectrogram directly. Smoothing allows retaining signal only where it is present across multiple frequencies e.g. where there is structure. 
Similar to Noisereduce, the noise threshold can be non-stationary and vary over time and adapt to periods of time with varying noise level. The code to run this method can be found here \url{https://github.com/RolnickLab/bioacoustics_denoising.git}

\begin{algorithm}[hbt!]
   \caption{2-Dimensional spectral gating}
   \label{alg:augmented_nr}
\begin{algorithmic}
   \STATE {\bfseries (a) Input:} Recording $X_{noise}$, z-score for noise thresholding $z_{th}$, Gaussian filter 2D size $(K, W)$
   \STATE {\bfseries (b) STFT:} Compute Short-Time Fourier Transform (STFT) of $X_{noise}$ and obtain a time series $S_n$ on each channel or frequency band $n$ 
   \STATE {\bfseries (c) Gaussian filter:} Run a Gaussian filter over $S$ to get a smoothed spectrogram $\tilde{S}$
   \STATE {\bfseries (d) Noise estimate:} Compute spectral statistics mean and standard deviation ($\mu_n, \sigma_n$) over each channel $S_n$
   \STATE {\bfseries (e) Masking:} Compute mask $M_\tau$ to keep values of $\tilde{S}_n$ above threshold $\tau_n = \mu_n + \sigma_n$
   \STATE {\bfseries (f) Smoothing:} Smooth the mask ($M_\text{smooth}$) with a filter over frequency and time.
   \STATE {\bfseries (g) Apply mask:} Apply the mask $M_\text{smooth}$ to the STFT of the signal ($S_X$) to produce the masked STFT ($S_M$)
   \STATE {\bfseries (h) Invert STFT:} Invert the masked STFT ($S_M$) back into the time-domain ($X_\text{denoised}$).
   \STATE {\bfseries (h) Output:} Denoised recording $X_\text{denoised}$
\end{algorithmic}
\end{algorithm}

\vspace{-0.3em}

\section{Experiments}\label{sec:experiments}

\subsection{Datasets}\label{sec:dataset}

We evaluate 2D-SG on the ``birdsong NOIZEUS'' dataset \cite{noizeus_birdsong}, a benchmark dataset modeled after NOIZEUS \cite{noizeus} methodology and structure. It consists of 70 recordings of 40 seconds, from 14 European starlings, recorded in an acoustically isolated chamber. These clean recordings are then added with noise from 8 soundscape categories from the ``Soundscapes from around the world'' dataset from Xeno Canto \cite{xeno_canto}, at four SNR levels: 0, 5, 10, and 15 dB. This simulates realistic conditions and allows to perform a quantitative evaluation by comparing the denoised signal with the clean recordings. 

We additionally evaluate 2D-SG qualitatively on two real datasets, shown in Fig. \ref{fig:results} (b, c). The first one consists of recordings of sperm whales in the Mediterranean sea. Sperm whales emit clicks, concentrated sound bursts, that are used for echolocation and social communication. Reliable detection of clicks is crucial for studying sperm whale behavior 
\cite{spermwhale2, spermwhale3}. The second dataset consists of recordings of European pied flycatchers \cite{LAMPE1994869}, songbirds with complex songs and vocal repertoire. Due to the complexity of their repertoire, it is important to have a clean, denoised vocalizations to study the differences between individuals and populations \cite{pied_flycatchers}. 

\subsection{Evaluation}\label{sec:evaluation}

Following the procedure in \cite{noise_reduce}, we compare 2D-SG with several denoising methods: Noisereduce \cite{noise_reduce}, the Wiener filter \cite{wiener1949}, and the Savitzky-Golay filter \cite{savitzky_golay}. Each method is evaluated using the following metrics: Segmental SNR (SegSNR \cite{seg_snr}), Scale-Invariant Signal-to-Distortion Ratio (SI-SDR, \cite{SI_SDR}), Log-Spectral Distance (LSD, \cite{lsd_metric}) and Mel Cepstral Distance (MCD, \cite{mcd_metric}). Seg-SNR evaluates the signal-to-noise ratio e.g. whether noise is effectively removed, and SI-SDR the quality of the reconstructed signal e.g. whether the reconstructed signal differs from the clean signal. 
The LSD evaluates reconstruction of the full spectrum while MCD focuses on the spectral envelope and ignores high frequencies. We estimate statistical significance between the results obtained using 2D-SG and Noisereduce by running a two-sample independent t-test at a 95\% confidence level.

All methods rely on one crucial hyperparameter, which controls the SNR and has the most effect on the denoising performance: 
the noise threshold (or z-score) for Noisereduce and 2D-SG, the degree of the polynomial for Savitzky-Golay, and the window size for the Wiener filter. To match how bioacousticians denoise in practice e.g. manually tune the main parameter of the denoising method for each recording, we tune this parameter by picking a random recording (``B335'' recording from Birdsong noizeus) and selecting the parameter that maximizes the SI-SDR for each SNR level and noise type. We then use this parameter to run and evaluate the methods on all other recordings. 

\begin{figure*}[htbp]
    \centering
    \includegraphics[width=\textwidth]{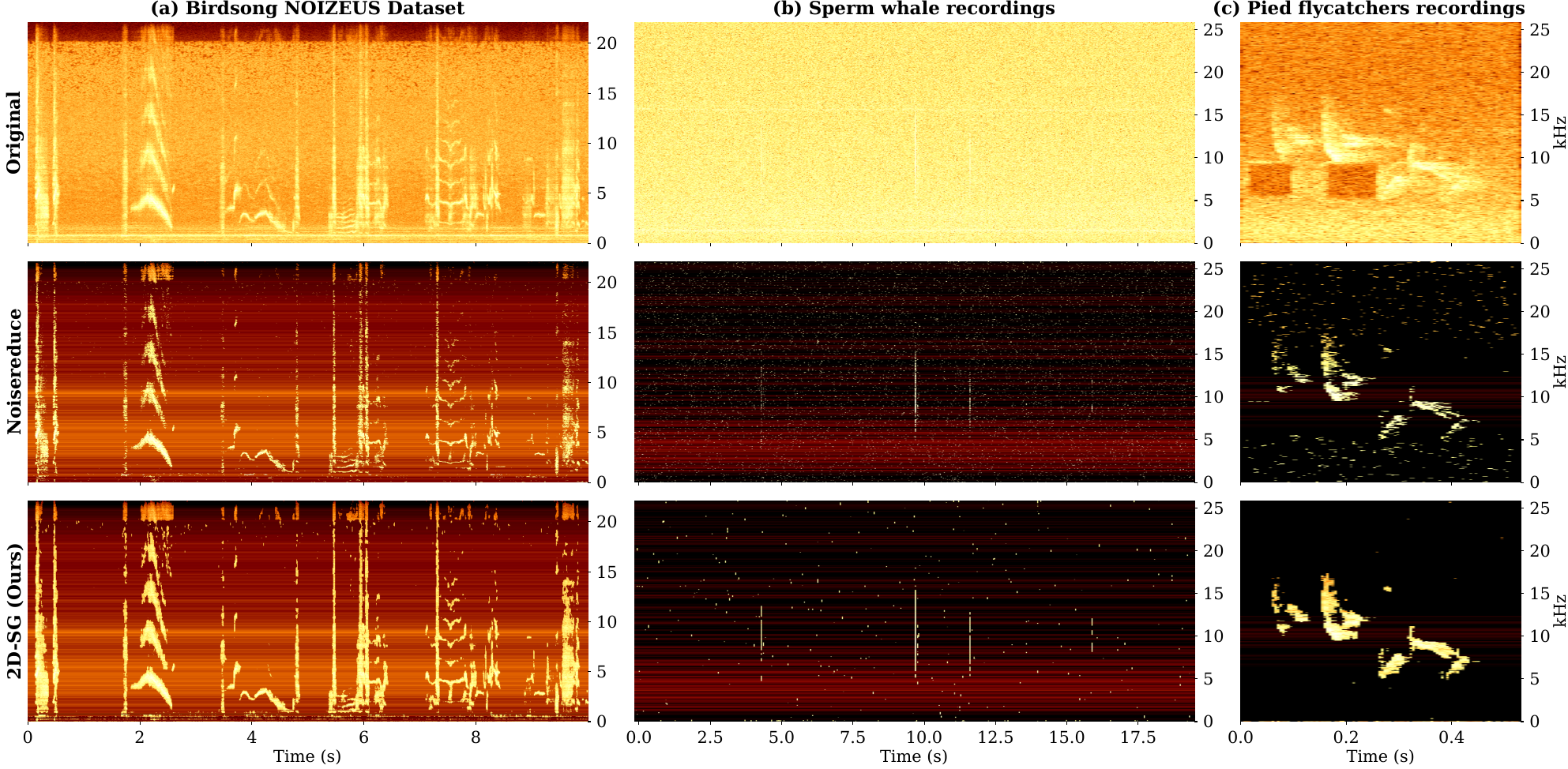} 
    \caption{\textbf{2-dimensional spectral gating qualitatively outperforms Noisereduce on three different recordings.} We show for (a) the B335 recording with ``town'' noise from the NOIZEUS dataset, (b) a sperm whale recording and (c) a European pied flycatcher recording: the raw spectrogram (upper), the spectrogram after having denoised with Noisereduce (middle) and the spectrogram with the 2-dimensional spectral gating (2D-SG, bottom). Colorbars are shared across each column.}
    \label{fig:results}
\end{figure*}

We show in Fig. \ref{fig:results} the raw spectrogram, as well as the spectrograms denoised with NoiseReduce and 2D-SG. For the sperm whale and pied flycatchers recordings, we tuned the noise threshold z-score parameter manually to obtain the visually best spectrograms. This procedure matches how acousticians denoise their recordings in practice and does not add processing time to their workflow, as the tuned parameter can then be used across the whole dataset. The final parameters used were, for Noisereduce, a z-score of 1.5 for all datasets. For 2D-SG, we used a z-score of 0.75 for the birdsong NOIZEUS dataset and the sperm whale recordings, and of 1 for the pied flycatchers dataset. The size of the Gaussian filter window used for all experiments in this paper was (0.25, 5) in the time-frequency domain.

\section{Results}\label{sec:results}

\begin{table}[hbt!]
\label{table:results}
\setlength{\tabcolsep}{3pt}
\resizebox{0.48\textwidth}{!}{
\begin{tabular}{c|c|cccc}
\hline
& Method & Seg-SNR ($\uparrow$) & SI-SDR ($\uparrow$) & MCD ($\downarrow$) & LSD ($\downarrow$)  \\
\hline
 \multirow{4}{*}{\rotatebox[origin=c]{90}{SNR 15}}
& Ours & \textbf{2.36}$^*$ & \textbf{11.94}$^*$ & \textbf{12.33}$^*$ & \textbf{4.75}$^*$ \\
& Noisereduce & 1.48 & 11.65 & 14.29 & 5.01 \\
& Wiener & 1.61 & 10.87 & 17.93 & 5.33 \\
& Savitzky-Golay & -0.77 & 9.09 & 35.46 & 6.80 \\
\hline
 \multirow{4}{*}{\rotatebox[origin=c]{90}{SNR 10}}
& Ours & \textbf{0.93}$^*$ & \textbf{8.55}$^*$ & \textbf{18.56} & \textbf{5.27}$^*$ \\
& Noisereduce & $-0.11$ & 8.13 & 20.13 & 5.79  \\
& Wiener & $-1.09$ & 6.07 & 33.76 & 6.84 \\
& Savitzky-Golay & $-3.58$ & 4.17 & 65.19 & 9.01 \\
\hline
 \multirow{4}{*}{\rotatebox[origin=c]{90}{SNR 5}}
& Ours & \textbf{0.04}$^*$ & \textbf{4.98} & \textbf{22.65}$^*$ & \textbf{6.33}$^*$ \\
& Noisereduce & $-0.83$ & 4.61 & 24.90 & 6.54 \\
& Wiener & $-3.50$ & 1.16 & 59.42 & 8.67 \\
& Savitzky-Golay & $-5.71$ & $-0.79$ & 108.72 & 11.30 \\
\hline
 \multirow{4}{*}{\rotatebox[origin=c]{90}{SNR 0}}
& Ours & \textbf{-0.58}$^*$ & 1.13 & \textbf{26.89} & \textbf{7.05} \\
& Noisereduce & $-0.97$ & \textbf{1.16} & 27.28 & 7.24 \\
& Wiener & $-5.55$ & $-3.82$ & 97.47 & 10.80 \\
& Savitzky-Golay & $-7.42$ & $-5.62$ & 166.65 & 14.17 \\
\end{tabular}
}
\caption{\textbf{2-dimensional spectral gating outperforms other methods on the birdsong NOIZEUS dataset.} We report four denoising metrics (from left to right, Seg-SNR, SI-SDR, MCD and LSD) for all four methods considered, on all recordings of the NOIZEUS dataset separated by SNR. 
* indicates statistical significance between Ours and the Noisereduce result estimated using a two-sample independent t-test at a 95\% confidence level.
}
\label{tab:results}
\end{table}

\subsection{Birdsong NOIZEUS dataset}

We report in table \ref{table:results} the mean metric e.g. averaged over all recordings of the birdsong Noizeus dataset per SNR level, and statistical significance of the difference between 2D-SG and Noisereduce. Our 2-dimensional spectral gating method outperforms all other evaluated methods on all 4 metrics, except the SI-SDR of SNR 0 recordings where the difference is not statistically significant.   
The increase in segmental SNR is statistically significant over NoiseReduce, and the increase in SI-SDR is statistically significant for SNR 10 and 15. The LSD and MCD are both lower for 2D-SG, for all SNR levels, with statistical significance for MCD of SNR 5 and 15 recordings, and LSD of SNR 5, 10 and 15 recordings.

An example birdsong NOIZEUS recording is shown in Fig. \ref{fig:results} (a). The top spectrogram shows the spectrogram of the noisy recording ``B335'' with noise type ``town''. Qualitatively, we see that 2D-SG (bottom spectrogram) removes a bit more noise than Noisereduce, especially between frequencies 15 and 20kHz, and better preserves the signal. For example, less signal is removed above frequency 20kHz, or across all frequencies around second 3.5. 

\subsection{Qualitative analysis on real-world recordings}

The middle panel of Fig. \ref{fig:results} shows results for Noisereduce and 2D-SG on the sperm whale recording. 2-dimensional spectral gating visually removes more noise, and also better preserve the signal, as a fourth click around second 16 becomes very apparent. This is promising and shows potential for accurate automatic detection of clicks in sperm whale recordings, which are essential for analyzing sperm whale communication. 2-dimensional spectral gating is particularly suited for such recordings since the clicks span many frequencies at once.  
The right panel shows results on a pied flycatcher recording, here corresponding to one song. 2-dimensional spectral gating again qualitatively outperforms Noisereduce here, as it both removes more noise and better preserves the signal, potentially allowing better studies of pied flycatchers' repertoire and communication.

\section{Discussion}\label{sec:discussion}

Although simple and efficient, this method requires one extra hyperparameter, the size of the Gaussian filter window, compared to Noisereduce. While this hyperparameter is intuitive as it should reflect the shape of the signal, default parameters might not yield optimal results on all vocalization types and 2D-SG might require manual tuning of this parameter as well as the z-score threshold. The method can also be extended, similarly as Noisereduce, to use an adaptive noise threshold. In presence of labeled data, the method will likely not outperform supervised machine learning based denoising methods, especially with low SNR data. However, 2D-SG does not require any machine learning expertise and can be run out-of-the-box without any training.  

The method shows very promising results for denoising but we haven't yet evaluated the improvements in downstream tasks, such as detection or sound event classification \cite{teng2025identifyingbirdsongsyllableslabelled}. To make the method available and easily applicable, we hope to integrate this method in audio processing softwares such as Audacity. Spectral gating methods such as Noisereduce are already implemented, and 2-dimensional spectral gating could be an option for processing acoustic recordings. 

\vfill\pagebreak



\section{Acknowledgments}
We thank David Rolnick, David Wheatcroft, Pierre Cauchy, Sam Lapp and members of the Kitzes lab for insightful discussions. This research was supported in part by the Canada CIFAR AI Chairs program and the Global Center on AI and Biodiversity Change (NSF OISE-2330423 and NSERC 585136). J.B is funded by the Fonds de recherche du Québec–
Nature et technologies (FRQNT; \url{https://doi.org/10.69777/2004813}).

\bibliographystyle{IEEEbib}
\bibliography{references}

\begin{thebibliography}{10}

\bibitem{PENAR2020100847}
W.~Penar, A.~Magiera, and C.~Klocek,
\newblock ``Applications of bioacoustics in animal ecology,''
\newblock {\em Ecological Complexity}, vol. 43, 2020.

\bibitem{xeno_canto}
``Xeno-canto - bird sounds from around the world.,''
\newblock {\em Xeno-canto Foundation for Nature Sounds.}, 2025.

\bibitem{noise_challenges}
R.~Gibb, E.~Browning, P.~Glover-Kapfer, and K.~E. Jones,
\newblock ``Emerging opportunities and challenges for passive acoustics in ecological assessment and monitoring,''
\newblock {\em Methods in Ecology and Evolution}, vol. 10, 2019.

\bibitem{background_noise_ml}
R.~Marshall-Hawkes, S.~Gillings, M.~W. Wilson, A.~S. Wetherhill, L.~V. Dicks, and A.~Ashton-Butt,
\newblock ``Estimating how site-level differences in acoustic environments affect species detection by machine learning models,''
\newblock {\em bioRxiv}, 2025.

\bibitem{bandpass_filter}
J.~M. Blackledget,
\newblock ``Chapter 4 - the fourier transform,''
\newblock in {\em Digital Signal Processing (Second Edition)}. Woodhead Publishing, 2006.

\bibitem{wiener1949}
N.~Wiener,
\newblock {\em Extrapolation, interpolation, and smoothing of stationary time series: With engineering applications},
\newblock MIT Press, 1949.

\bibitem{savitzky_golay}
A.~Savitzky and M.~Golay,
\newblock ``Smoothing and differentiation of data by simplified least squares procedures,''
\newblock {\em Analytical Chemistry}, vol. 36, 1964.

\bibitem{noise_reduce}
T.~Sainburg and A.~Zorea,
\newblock ``Domain general noise reduction for time series signals with noisereduce,''
\newblock {\em Scientific Reports}, vol. 15, 2025.

\bibitem{AutoencoderDenoiser1}
R.~Sinha and P.~Rajan,
\newblock ``A deep autoencoder approach to bird call enhancement,''
\newblock in {\em 2018 IEEE 13th International Conference on Industrial and Information Systems (ICIIS)}, 2018.

\bibitem{amine_denoiser}
A.~Razig, Y.~Soulaymani, L.~Benabbou, and P.~Cauchy,
\newblock ``Multi-representation attention framework for underwater bioacoustic denoising and recognition,'' 2025.

\bibitem{noise2noise}
M.~M. Kashyap, A.~Tambwekar, K.~Manohara, and S.~Natarajan,
\newblock ``Speech denoising without clean training data: A noise2noise approach,''
\newblock in {\em Interspeech 2021}. 2021, ISCA.

\bibitem{jerri1998gibbs}
A.~J. Jerri,
\newblock {\em The Gibbs Phenomenon in Fourier Analysis, Splines and Wavelet Approximations},
\newblock Springer Science \& Business Media, 1998.

\bibitem{noizeus_birdsong}
T.~Sainburg and A.~Zorea,
\newblock ``Birdsong noizeus: Bioacoustics noise reduction benchmark dataset,'' 2024.

\bibitem{noizeus}
Y.~Hu and P.~C. Loizou,
\newblock ``Subjective comparison and evaluation of speech enhancement algorithms,''
\newblock {\em Speech Communication}, vol. 49, 2007,
\newblock Speech Enhancement.

\bibitem{spermwhale2}
P.~Sharma, S.~Gero, R.~Payne, D.~F. Gruber, D.~Rus, A.~Torralba, and J.~Andreas,
\newblock ``Contextual and combinatorial structure in sperm whale vocalisations,''
\newblock {\em Nature Communications}, vol. 15, 2024.

\bibitem{spermwhale3}
J.~Andreas, G.~Beguš, M.~M. Bronstein, R.~Diamant, D.~Delaney, S.~Gero, and S.~{Goldwasser et al.},
\newblock ``Toward understanding the communication in sperm whales,''
\newblock {\em iScience}, vol. 25, 2022.

\bibitem{LAMPE1994869}
H.~M. Lampe and Y.~O. Espmark,
\newblock ``Song structure reflects male quality in pied flycatchers, ficedula hypoleuca,''
\newblock {\em Animal Behaviour}, vol. 47, 1994.

\bibitem{pied_flycatchers}
M.~Gallego-Abenza, F.~H. Kraft, L.~Ma, S.~Rajan, and D.~Wheatcroft,
\newblock ``Responses in adult pied flycatcher males depend on playback song similarity to local population,''
\newblock {\em Behavioral Ecology}, vol. 36, 2025.

\bibitem{seg_snr}
P.~Mermelstein,
\newblock ``Evaluation of a segmental snr measure as an indicator of the quality of adpcm coded speech,''
\newblock {\em The Journal of the Acoustical Society of America}, vol. 66, 1979.

\bibitem{SI_SDR}
J.~Le~Roux, S.~Wisdom, H.~Erdogan, and J.~R. Hershey,
\newblock ``Sdr – half-baked or well done?,''
\newblock in {\em 2019 IEEE International Conference on Acoustics, Speech and Signal Processing (ICASSP)}, 2019.

\bibitem{lsd_metric}
A.~Gray and J.~Markel,
\newblock ``Distance measures for speech processing,''
\newblock {\em IEEE Transactions on Acoustics, Speech, and Signal Processing}, vol. 24, 1976.

\bibitem{mcd_metric}
R.~{Kubichek},
\newblock ``{Mel-cepstral distance measure for objective speech quality assessment},''
\newblock in {\em Proceedings of IEEE Pacific Rim Conference on Communications Computers and Signal Processing}, 1993, vol.~1.

\bibitem{teng2025identifyingbirdsongsyllableslabelled}
M.~Teng, J.~Boussard, D.~Rolnick, and H.~Larochelle,
\newblock ``Identifying birdsong syllables without labelled data,''
\newblock in {\em 2026 IEEE International Conference on Acoustics, Speech and Signal Processing (ICASSP)}, 2026.

\end{thebibliography}

\end{document}